\documentstyle[aps,prl,preprint,tighten]{revtex}

\begin{document}
\draft
\title{Spin 1 inversion: a Majorana tensor force for deuteron alpha scattering}
\author{S.G. Cooper$^{\dag}$, V.I. Kukulin$^{\ddag}$, R.S. Mackintosh$^{\dag}$
 and  V.N. Pomerantsev$^{\ddag}$}
\address{$^{\dag}$Physics Department, The Open University, Milton Keynes,
 MK7 6AA, U.K.\\$^{\ddag}$Institute of Nuclear Physics, Moscow State 
University, Moscow 119899, Russia. }

\date{\today}
\maketitle
\begin{abstract}
We demonstrate, for the first time, successful S-matrix to potential inversion
for spin one projectiles with non-diagonal $S^j_{ll'}$ yielding a $T_{\rm R}$ 
interaction. The method is a generalization of the iterative-perturbative, IP, method. 
We present a test
case indicating the degree of uniqueness of the potential. The method is adapted,
using established procedures, into direct observable to potential inversion, 
fitting $\sigma$, ${\rm i}T_{11}$, $T_{20}$, $T_{21}$ and $T_{22}$ for d + alpha
scattering over a range of energies near 10 MeV. The $T_{\rm R}$ interaction
which we find is very different from that proposed elsewhere, both real and
imaginary parts being very different for odd and even parity channels.
\end{abstract}

\pagebreak

\setlength{\parindent}{0.3 in}

The inverse scattering problem, i.e.\ the problem of determining a potential 
from the scattering matrix,
has not until now been solved for the scattering of spin one projectiles. 
Spin one inversion is substantially different from that for spin zero
or spin  half because, even for a spin-zero target, the elastic scattering 
`$S$-matrix' is no longer a complex number but a matrix. 
When spin-1 projectiles scatter from a spin-zero target, 
the conserved quantities are total angular
momentum $j$ and parity $\pi$. When $\pi = (-1)^{j+1}$,  two  values of orbital angular 
momentum, $l$ and $ l'$, are, in general, coupled by a tensor interaction.  
Calculating the
$S$-matrix $S^j_{ll'}$ therefore requires a coupled channel calculation and 
inversion from $S^j_{ll'}$ to a potential containing a tensor term  requires 
a coupled channel inversion.  Here we demonstrate a spin-one
inverse scattering procedure. It is incorporated into
a direct data-to-potential inversion procedure which is used to analyse  
$^2$H -- $^4$He scattering data at around 10 MeV. We find evidence for a 
substantial parity dependent tensor interaction.

Coupled channel inversion  is here carried out using a generalization
of  the iterative perturbative (IP)  inversion 
procedure\cite{early,ketal1,candm}, which is successful
for $S_{lj} \rightarrow V(r) + {\bf l \cdot s} V_{\rm so}(r)$ inversion for
spin half projectiles. Features of the IP method
(e.g.\ the ability to handle a range of energies simultaneously 
and to include Majorana terms
for all potential components) all apply.

We seek a potential with many components, each component being labelled with 
index $p$ identifying central,
spin-orbit or tensor terms, each real or imaginary. The number of components
further doubles where, as essential for light nuclei, parity dependence is permitted.
We assume that the tensor force is of the form  
$\hat{V}^{(t)}=T_{\rm R} V^{(t)}(r) \equiv ({\bf (s\cdot \hat{r})^2} -2/3)V^{(t)}(r)$
\cite{satchler}, where $\hat{V}^{(t)}$ is assumed to  have both Wigner 
and Majorana terms.

The IP method commences from a `starting reference potential', SRP and
the potential is corrected  iteratively. At each iteration,
all components, indexed by $p$, are modified  by adding a superposition 
$\sum \alpha^{(p)}_n v_n(r)$  where functions $v_n(r)$ are members of
an `inversion basis' (which, if required, can be chosen differently for
different $p$.)  The amplitudes $\alpha^{(p)}_n$ 
are determined at each iteration from linear equations, based on an SVD algorithm,
which successively reduce $\sigma^2 = \sum |S^{\rm t}_k - S^{\rm c}_k|^2$. For
each partial wave $k$, $S^{\rm t}_k$ is the `target' $S$-matrix and $S^{\rm c}_k$
is for the potential at the current iteration. Here label $k$ is a single index which
identifies  non-diagonal elements of $S^j_{ll'}$ as well as the energy $E_i$ when
$S^j_{ll'}(E_i)$ for energies $E_i$ are simultaneously inverted.
The physical basis of the equations leading to $ \alpha^{(p)}_n$ is the 
the linear response
of $S_k$ to small perturbations in the potential\cite{early,ketal1,candm}.
This result applies to the non-diagonal as well as diagonal
terms. For any given set of conserved quantum numbers, certain
channels will be coupled by the nucleus-nucleus interaction and
we use labels $\kappa, \lambda, \mu, \nu$ for these channels. Thus the matrix element of
the nucleus-nucleus interaction $V$ between the wavefunctions for channels $\kappa$ and
$\lambda$, corresponding to integrating
over all coordinates but $r$, will be written $V_{\kappa\lambda}(r)$.
The increment $\Delta S_{\kappa\lambda}$ in the non-diagonal S-matrix which is due to
a small perturbation $\Delta V_{\kappa\lambda}(r)$,  is\begin{equation}
 \Delta S_{\kappa\lambda} = \frac{{\rm i}\mu}{\hbar^2 k} 
 \sum_{\mu\nu}\int_0^{\infty}\psi_{\mu\kappa}(r) \Delta V_{\mu\nu}(r)
\psi_{\nu\lambda} {\rm d}r \label{integ}\end{equation} 
where $\psi_{\nu\kappa}$ is the $\nu$th channel (first index) component 
of that coupled channel solution for the unperturbed non-diagonal potential
for which there is in-going flux in channel $\kappa$ (second index) only.
The normalisation is
$\psi_{\nu\kappa} \rightarrow \delta_{\kappa\nu} I_{l_{\kappa}} - S_{\nu\kappa} 
O_{l_{\nu}}$ where
$I_l$ and $O_l$ are incoming and outgoing Coulomb wavefunctions for
orbital angular momentum $l$; there is
no complex conjugation in the integral. Starting from Eq.\ref{integ}, spin-one
inversion becomes a straightforward generalisation of the procedure described
in Refs.~\cite{candm} and is now implemented in the code IMAGO~\cite{imago}.
 
{\bf Testing spin-1 $S \rightarrow V$ inversion.} 
We established that in forward (rather than inverse) 
mode, a given $T_{\rm R}$ potential led to the same $S^j_{ll'}$ and observables
(including tensor analysing powers) as the standard deuteron scattering code
DDTP~\cite{ddtp}. A convenient property of the IP inversion method is the fact that 
when a sequence of iterations converges to small $\sigma^2$, as defined above, 
we can be sure that we have found a potential which reproduces the set of
target $S^j_{ll'}$ to a precision which corresponds to that value of $\sigma^2$.  
It follows that the forward-mode test verifies that 
we have found a correctly defined potential. 

We have also tested the prescription for spin-1 $S \rightarrow V$ inversion 
using a known potential to define the target $S^j_{ll'}$.
One set of results, for 20.2 MeV deuterons scattering 
from $^4$He is shown in Fig.~1. The target potential~\cite{frickprl} (solid line) 
has a complex central term and real ${\bf l \cdot s}$ and
$\hat{V}^{(t)}$ terms but no Majorana components.
This potential reproduces the tensor analysing powers but with a 
$T_{\rm R}$ interaction much stronger than folding models~\cite{kaprc8} predict. 
The two inversion solutions shown in Fig.~1 both arise from a starting potential
(SRP) with only one non-zero component, a real central Woods-Saxon potential 60
MeV deep. Although few partial waves contribute significantly to the inversion,
only 3-4 iterations were necessary to give solution A  (dotted) and
solution B (dot-dashed), which reproduced all four volume integrals to better than
1 \%, required not many more.Defects in the tensor and spin-orbit terms 
for $r\sim 0$ reflect the fact that both  ${\bf l \cdot s}$ and $ T_{\rm R}$
have zero diagonal matrix elements for $l=0$. A test case involving parity 
dependence would
necessitate a smaller inversion basis, and hence reproduce each component 
with less accuracy, since $S^j_{ll'}$ for the same range of $j$ must determine
twice as many components; this is relevant to the case described below. 

{\bf Direct data-to-potential inversion.}
The preferred method of applying inversion methods
to experimental scattering data is to convolute the phase shift fitting  
with the  $S \rightarrow V$ inversion to form a direct  data-to-potential inversion
procedure. This can be coded in a single data fitting program. 
The two step procedure in which one first
fits the data with $S^j_{ll'}$ and then  subsequently 
inverts $S^j_{ll'}$ does not guarantee that the $S^j_{ll'}$
for different energies are smoothly related like $S^j_{ll'}$
derived from a single potential. The direct inversion
formalism has been described and applied~\cite{convol1,convol2} to cases 
where there is no channel coupling and it
 carries over to the present case with no change in basic principle. 
 It has been implemented in code IMAGO~\cite{imago} and the convolution
was thoroughly tested through explicit checking of all the
derivative terms which are required ~\cite{convol1,convol2}. 

{\bf Application to deuteron scattering.} 
We have analysed the scattering 
of tensor polarised deuterons from $^4$He over a series of energies near 10 MeV
using direct data-to-potential inversion. 
This case is of particular interest both because it should become feasible to
do quite realistic calculations using RGM or similar methods and also
because the deuteron is the archetypal easily polarisable halo nucleus. 
From the data set for d -- alpha scattering tabulated
in Ref.~\cite{kuznetsova}, we have selected data of Jenny 
{\em et al\/}\cite{jenny} which is of high quality and covers a wide angular range.  
We fit the following observables,
$\sigma$, ${\rm i}T_{11}$, $T_{20}$, $T_{21}$ and $T_{22}$,
for the energies, 8, 9, 10, 11, 12 and 13 MeV. This range includes broad 2$^+$
resonances and a region of strong mixing between the 1$^+$ channels.  
Ref~\cite{prl58}  applies direct data-to-potential IP inversion
 to data from the same source but without fitting $T_{2i}$.

We exploit a property of the IP method to fit
all of the above data, for all the stated energies, with  a single
potential. The potential is of fixed radial form but the overall strength of 
each of the imaginary components
is proportional to the energy above the inelastic threshold. 
To minimize the number of parameters  the real components are taken to be energy
independent.

It is well known that fits of $S$-matrix to data are inevitably subject to
discrete and continuous ambiguities and direct data-to-$V$ inversion is not 
exempt. The uncertainties arising at the $S \rightarrow V$ stage are less
important. Various strategies are possible, such as the inclusion
of {\em  a priori\/} information from theory~\cite{puri}. Here
we adopt the approach of using extremely small
inversion bases (two or three Gaussian functions) so that, in effect, we 
find the smoothest, minimally energy-dependent, potential
fitting the data at all six energies. We find that we can allow the
spin orbit terms  to be real and parity independent but the central and $T_{\rm R}$ 
tensor terms must be complex and  and parity dependent. Inversion studies
involving fits to both empirical data and RGM theory~\cite{prc54,emit}
show that interactions between light nuclei are parity dependent for energies up to
tens of MeV per nucleon. As in earlier work,
we present the real central term in the form of Wigner and Majorana components:
$V_{\rm W} + (-1)^l V_{\rm M}$. However, $\hat{V}^{(t)}$  and the imaginary terms
were treated somewhat differently:
the data was reproduced most efficiently with inversion bases covering different
radial ranges  for the even and odd parity components. In Fig.~3 below
we shall see that the odd parity  $\hat{V}^{(t)}$ terms cover a wider radial range.
This is consistent with theoretical arguments~\cite{fb98} 
for the existence of tensor interactions of quite different strengths and shapes for 
odd and even parity. The SRP
consisted only of real, Wigner, central and spin-orbit terms.
Convergence was very rapid and typically about 3 iterations were required.

Figs 2 and 3 show two alternative potentials fitting the data for the six energies. 
The energy dependent imaginary parts were evaluated at 10 MeV and Fig.~4 shows the
quality of  fit to the 10 MeV data. The fit to
the data at the ends of the energy range was poorer suggesting
that a somewhat less simple energy dependence is required.   
The  $\chi^2/N$ values of 8.59 and 8.08 are for the data at all six 
energies.\footnote{For simplicity in this initial study, we made 
no selection, filtration or normalisation of input data  as is usual
in phase shift analyses.}

Different components of the potential are determined to different 
degrees of uniqueness. The real central 
and spin-orbit Wigner potentials shown in Fig. 2 are well determined and 
consistent with predictions of RGM theory.
The imaginary central component is highly parity dependent and is
emissive  near the origin. 
All potentials giving a good fit to the data over this energy range have
such an emissive region near the nuclear centre.
Such emissive regions 
are commonly present~\cite{emit} in potentials found by inverting RGM $S$-matrix 
elements which manifestly do not break  unitarity. Emissive regions often appear 
in local potentials which represent exchange and channel coupling non-locality. 
In the present case, the unitarity limit is only broken,
to  a small degree, for $L=2$, $J=2$ for the  $\chi^2/N =  8.59$ and is not broken
for the better fit. 

The  $V^{(t)}(r)$  interaction shown in Fig. 3 departs markedly 
from previous phenomenology.
The even parity real and imaginary  terms are very strong
near $r=1$. This is significant for $r > $ about 1.0 fm; for  $r<1$
the even-parity  $V^{(t)}(r)$ interaction is ill-determined 
since the diagonal matrix element of $T_{\rm R}$  
is zero for $l<2$ 
so the comment concerning irregularities at the nuclear centre
in Fig. 1 applies here. The combination of strong, short-ranged
even-parity components and weaker, long ranged odd-parity components is precisely
the pattern predicted to arise from a deuteron exchange mechanism
discussed in Ref.~\cite{fb98}; see also Refs.~\cite{dubo98,kuk-new} which also
present a $\hat{V}^{(t)}$  interaction strongly peaked inside $r=2$ fm, 
but of the opposite sign. 

The present analysis firmly establishes that most terms in the d -- $^4$He
potential are  substantially parity dependent, casting doubt
on parity independent d -- $^4$He potentials, including
the potential of Frick {\em et al\/}~\cite{frickprl} with its remarkably large
$\hat{V}^{(t)}$  interaction.
RGM theory~\cite{puri,emit} predicts parity dependence for central
components, but the strong parity dependence in $\hat{V}^{(t)}$  suggests a new 
physical process.

In conclusion: we have demonstrated for the first time $S \rightarrow V$ inversion
for spin-1 projectiles leading to a tensor interaction which couples channels. 
This has  been incorporated into a direct data-to-potential inversion procedure
yielding a remarkable new kind of $T_{\rm R}$ interaction.
The general procedure could be applied to other cases with  channel spin one
such as p -- $^3$H scattering. Spin-1
$S \rightarrow V$ inversion  makes it possible to 
study theories  for dynamic polarization and exchange 
contributions to central and non-central forces for spin-1
projectiles.  The results can then be directly
related to phenomenology since, as we have shown here,
the inversion procedure can be made part of a powerful,
rapidly converging, phenomenological
method which, at low computational cost, fits multi-energy datasets with a 
single, energy dependent potential.

\section*{Acknowledgements}  V.I.K. is grateful to Willi
Gruebler for supplying the full tables of experimental data of the
Z\"urich group.  We are also grateful to the UK EPSRC for grant
GR/L22843 supporting S.G. Cooper, the Russian Foundation for Basic
Research (grant 97-02-17265) for financial assistance and to the Royal
Society (UK) for supporting a visit by V.I. Kukulin to
the UK. We thank Jeff Tostevin for sending us Goddard's deuteron scattering code
DDTP.

\setlength{\parindent}{0.0 in}

\newpage

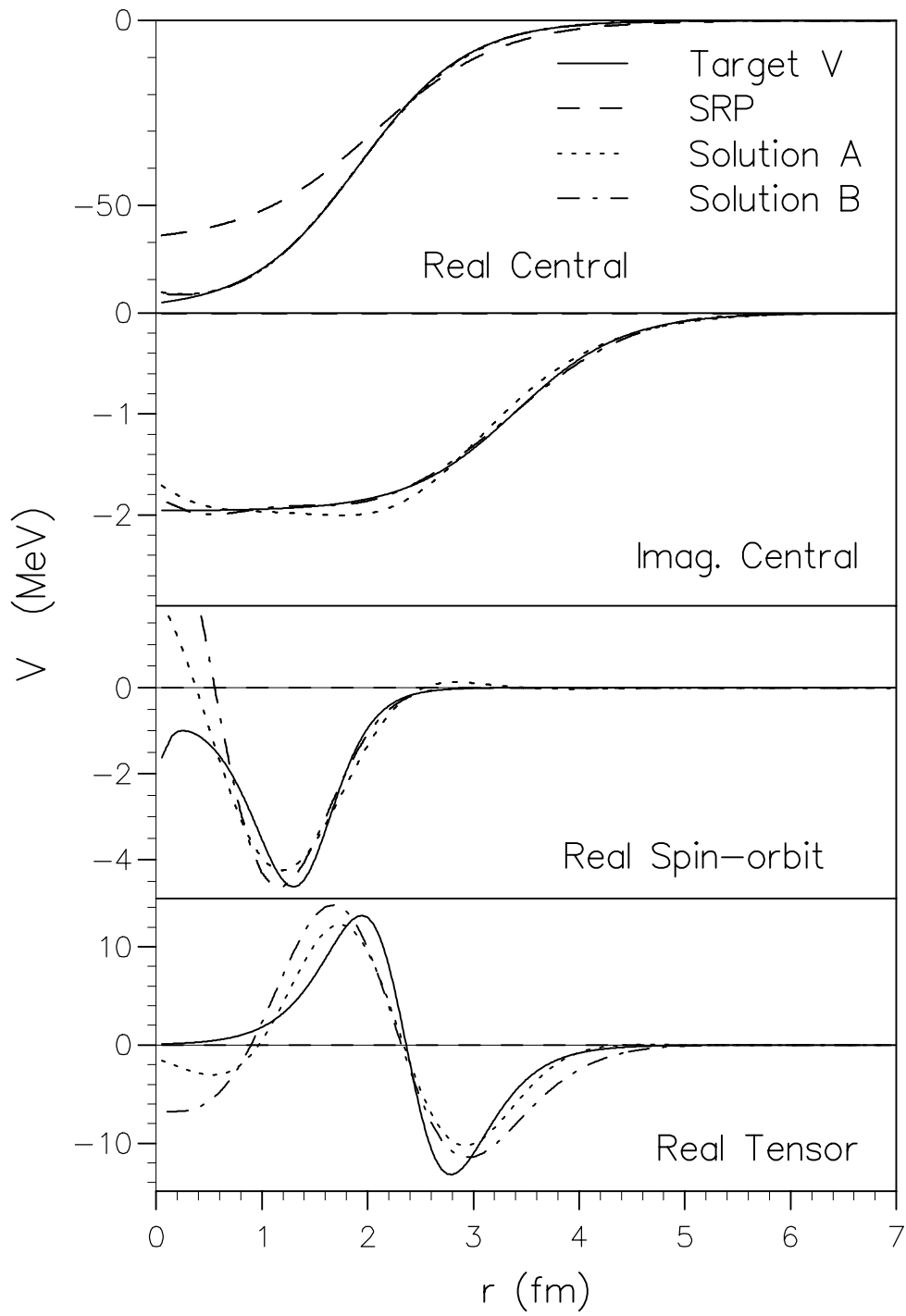
\begin{figure}
\caption{For deuterons scattering from $^4$He at 20.2 MeV, the full line represents
the potential of Frick {\em et al\/}, the long dashes represent the SRP
(non-zero only for the real central term), the dots represent inversion
solution A, and the dot-dashes represent solution B.}
\end{figure}

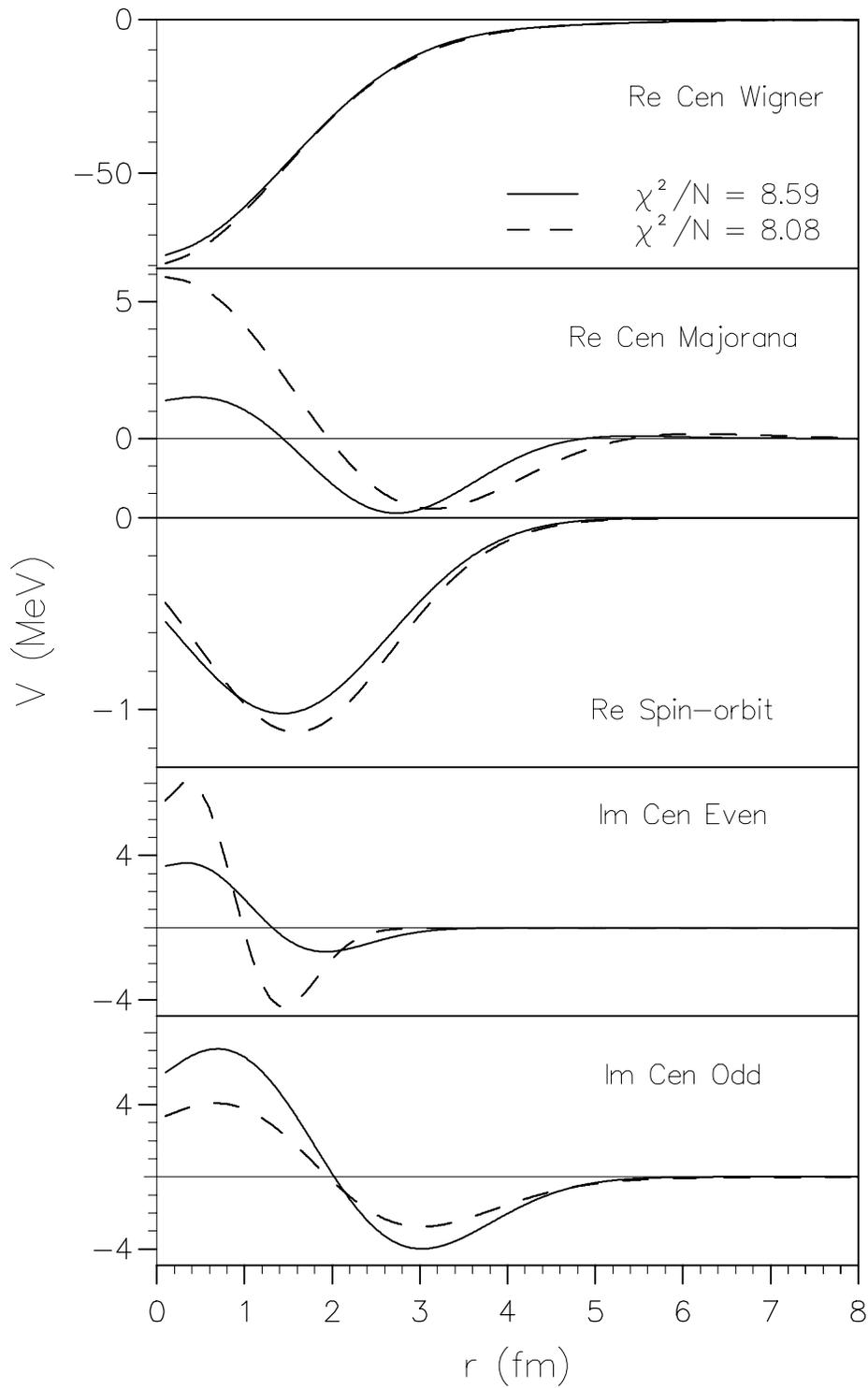
\begin{figure}
\caption{From top: Real, Wigner and Majorana, central components; real, Wigner  spin-orbit
component; imaginary,  central, even and odd parity components. All  were evaluated
at 10 MeV, and correspond to
inverted potentials with $\chi^2/N = 8.59$ (solid line) and $\chi^2/N = 8.01$
(dashed line).}
\end{figure}

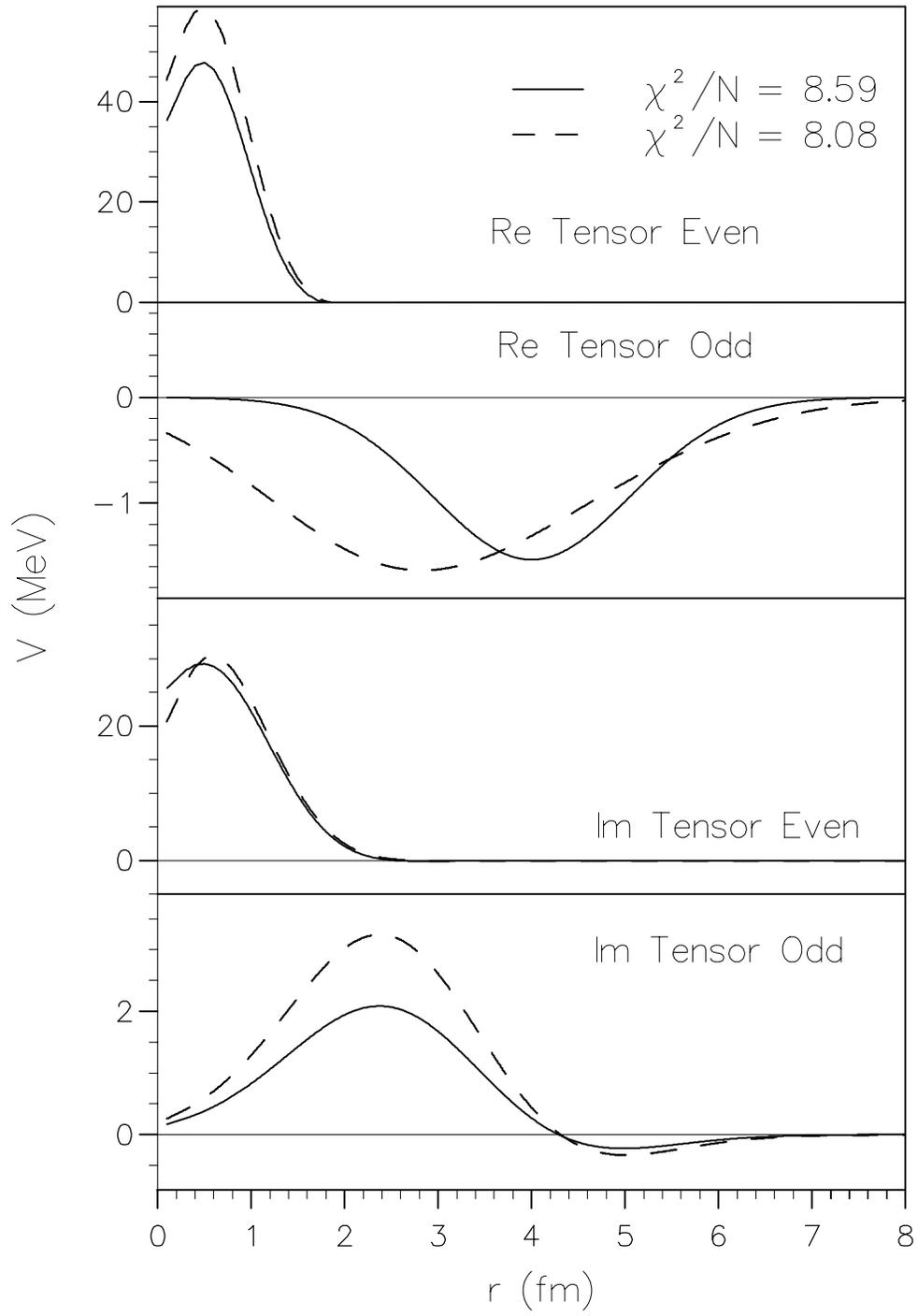
\begin{figure}
\caption{Tensor components of the potentials of
Figure 2, from top: Even and odd, real then even and odd  imaginary. 
Again, solid line is for $\chi^2/N = 8.59$ and dashed line is for $\chi^2/N = 8.01$.}
\end{figure}

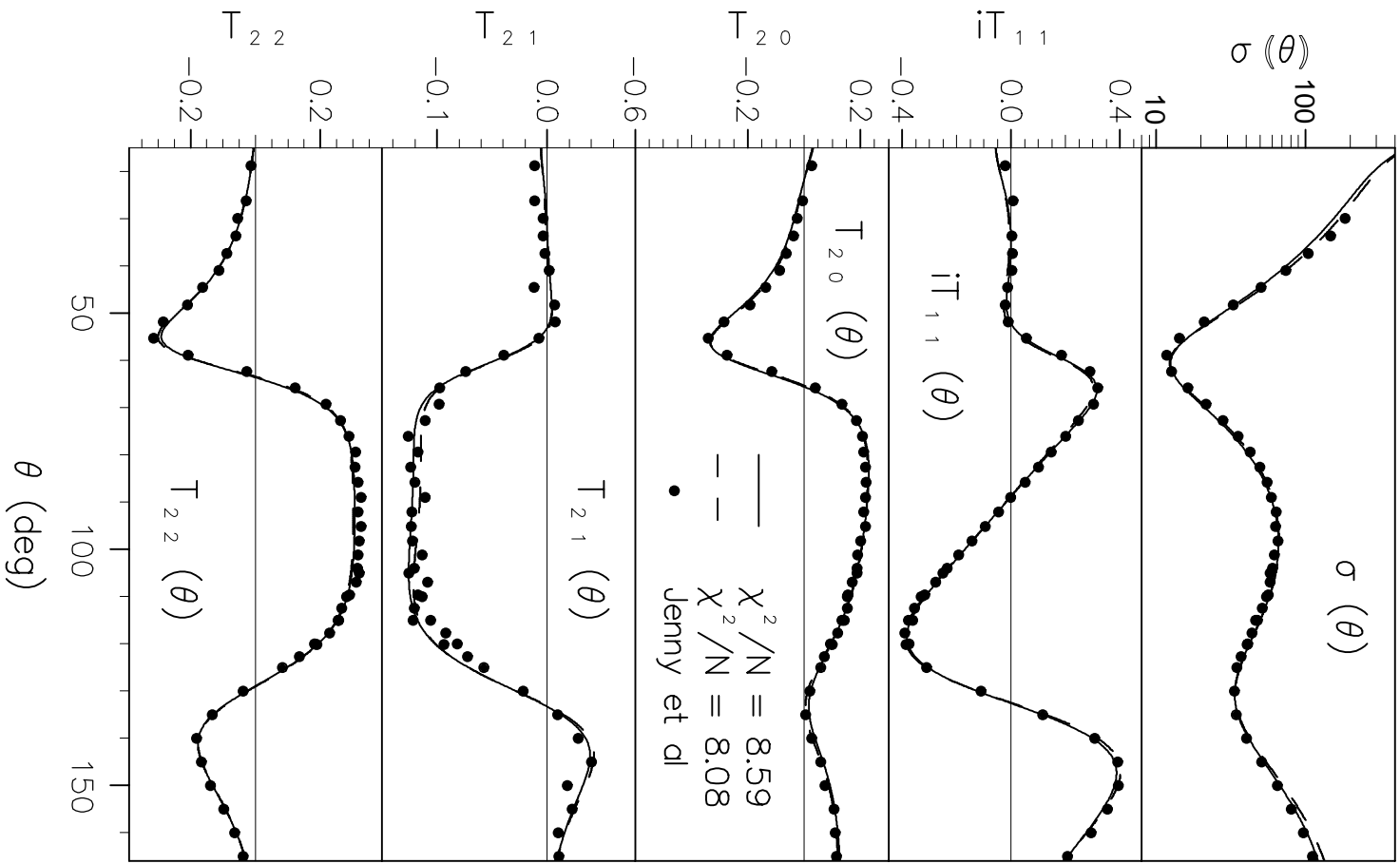
\begin{figure}
\caption{For deuterons scattering from $^4$He at a laboratory energy of 10 MeV,
$\sigma(\theta)$, i$T_{11}(\theta)$, $T_{20}(\theta)$, 
$T_{21}(\theta)$, $T_{22}(\theta)$ given by the potentials shown in
Figures 2 and 3, with $\chi^2/N = 8.59$ (solid line) and $\chi^2/N = 8.01$
(dashed line), compared with the data of Jenny {\em et al\/} (solid points).}
\end{figure}

\end{document}